\documentclass[aip,
rsi,%
 amsmath,amssymb,
 reprint,%
]{revtex4-1}

\usepackage{graphicx}
\usepackage{dcolumn}
\usepackage{bm}
\usepackage[version=3]{mhchem}
\usepackage{threeparttable}
\usepackage{color}
\usepackage{ulem}
\usepackage{xr}
\begin{document}

\title{A CCSD(T)-based permutationally invariant polynomial 4-body potential for water}
\date{\today}
\author{Apurba Nandi}
\email{apurba.nandi@emory.edu}
\affiliation{Department of Chemistry and Cherry L. Emerson Center for Scientific Computation, Emory University, Atlanta, Georgia 30322, U.S.A.}
\author{Chen Qu}
\affiliation{Department of Chemistry \& Biochemistry, University of Maryland, College Park, Maryland 20742, U.S.A.}
\author{Paul L. Houston}
\email{plh2@cornell.edu}
\affiliation{Department of Chemistry and Chemical Biology, Cornell University, Ithaca, New York
14853, U.S.A. and Department of Chemistry and Biochemistry, Georgia Institute of
Technology, Atlanta, Georgia 30332, U.S.A}
\author{Riccardo Conte}
\email{riccardo.conte1@unimi.it}
\affiliation{Dipartimento di Chimica, Universit\`{a} Degli Studi di Milano, via Golgi 19, 20133 Milano, Italy}
\author{Joel M. Bowman}
\email{jmbowma@emory.edu}
\affiliation{Department of Chemistry and Cherry L. Emerson Center for Scientific Computation, Emory University, Atlanta, Georgia 30322, U.S.A.}


\begin{abstract}
We report a permutationally invariant polynomial (PIP) potential energy surface for the water 4-body interaction.  This 12-atom  PES is a fit to 2119, symmetry-unique, CCSD(T)-F12a/haTZ (aug-cc-pVTZ basis for `O' atom and cc-pVTZ basis for `H' atom) 4-b interaction energies.  These come from low-level, direct-dynamics calculations,  tetramer fragments from an MD water simulation at 300 K, and from the water hexamer, heptamer, decamer, and 13-mer clusters.  The PIP basis is purified to ensure that the 4-b potential goes rigorously to zero in monomer+trimer and dimer+dimer dissociations for all possible such fragments. The 4-b energies of isomers of the hexamer calculated with the new surface are shown to be in better agreement with benchmark CCSD(T) results than those from the MB-pol potential. Other tests validate the high-fidelity of the PES.
\end{abstract}
\maketitle

\section{Introduction}
The many-body expansion (MBE) for non-covalent interaction is pervasive and has been applied in  many guises for water.  There are numerous studies of this expansion for water and many of these are discussed in a review article.\cite{mbreview} The general conclusion from numerical studies is that an accurate, i.e., to within a few percent, description can be obtained with the MBE truncated at  3-body interactions.  However, it has been known for many years that the 4-b interaction, while small, may not be negligible. As early as 2000, Xantheas concluded that for the hexamer the 4-b contribution to the total interaction energy varies from 1.4 to 4.4 percent, depending on the isomer.\cite{XANTHEAS2000}  In 2011 a many-body analysis of the lowest energy isomers for the hexamer, the prism and the cage, determined from CCSD(T)/aug-cc-pVTZ calculations that 4-b contributions to the interaction energy were -0.57 and -0.43 kcal/mol, respectively.\cite{KShexamer} This lowers the prism energy by about 49 cm$^{-1}$ more than the cage. Recently, Heidel and Xantheas reported a careful numerical examination (focusing on elimination of basis set superposition errors) of the MBE for interaction energies of water clusters, consisting of 7, 10, 13, 16 and 21 monomers.\cite{MBE20} Based on this and previous analyses, they concluded that the MBE can thus be safely (i.e., converged to less than 1 percent) truncated at the 4-body term and not at the 3-body level.

Given that the MBE truncated at the 3-body term does capture most of the interaction energy of water monomers, several \textit{ab initio} water potentials have been developed based on 1, 2 and 3-body terms. These are known by the acronyms WHBB,\cite{WHBB} HBB2-pol, MB-pol,\cite{paesani16} and CC-pol.\cite{ccflex23} The WHBB, HBB2-pol, and MB-pol potentials use the spectroscopically accurate 1-b potential,\cite{PS} and for the 2-b numerical fits, using permutationally invariant polynomials,\cite{Braams2009} of thousands of CCSD(T) energies.\cite{HBB1,HBB2,mbpol2b}  The 3-b potential in WHBB is a PIP fit to thousands of MP2 energies, whereas the one in the HBB2-pol and MB-pol potentials is based on fits to thousands of CCSD(T) energies. The CC-pol potential uses elaborate functional forms to represent the 2 and 3-b interactions and to fit them to many CCSD(T) energies.\cite{ccflex23} 

The WHBB potentials account for 4 and higher-body interactions by switching to TTM3-F potentials which provide a sophisticated treatment of long range electrostatics for an arbitrary number of monomers.\cite{TTM3F}  The HBB2-pol and MB-pol potentials  used a similar strategy to switch to a classical electrostatic potential in the long-range; however, TTM4-F\cite{TTM4} was used instead of TTM3-F, plus additional electrostatic terms. A critical assessment of the accuracy of the TTM3-F and TTM4-F potentials for 4-b energies of isomers of the water hexamer against direct CCSD(T)-F12b/VTZ was reported in 2015.\cite{medders15} While the TTM4-F potential is generally more accurate than the TTM3-F one, it has errors between around 0.1 and 0.35 kcal/mol.  These are fairly large fractions of the 4-b energy itself.  Indeed these errors account  substantially for the overall errors of the WHBB and MB-pol potentials for the interaction energies of hexamer isomers. In addition, these errors are potentially significant for rigorous studies of the relative energies of the prism and cage, which have been done with both the WHBB and MB-pol potentials and using rigorous treatments of the vibrational motion.\cite{wanghex} In the latest study using MB-pol, the cage was reported to be more stable at 0 K than the prism by roughly 0.1 kcal/mol, including a rigorous treatment of zero-point energy.\cite{Mandprismhex} This difference is within the range of errors in the hexamer 4-b interaction in the MB-pol potential and so the conclusion probably should be viewed with some caution.

Based on this and previous work, there is strong motivation to develop a full-dimensional PES for the 4-b water interaction potential.  This is a challenging 12-atom system that was beyond consideration during the time when WHBB, MB-pol, and CC-pol were developed. However, we have recently extended the PIP approach to describe PESs for 12-atom $N$-methyl acetamide\cite{QuBowman2019, NandiQuBowman2019} and 15-atom tropolone\cite{tropolone20} and acetylacetone.\cite{QuPCCP2020}  Because the long-range behavior is important for the  4-b interaction, we use a PIP basis that is purified\cite{purified13, purified14, purified15a} so that the 4-b PES goes rigorously to zero as any monomer (or dimer) separates from the other group of monomers. The CCSD(T)-F12a/haTZ method was selected for the calculations of electronic energies as it provides accurate results at an acceptable computational cost.

Below we give details of the generation of the database of electronic energies and standard tests of the precision of the 4-b PES are given.  Then the new PES is tested for the 4-b energies of 8 isomers of the water hexamer, and the total 4-b energies of a number of larger water clusters. The long-range behavior is examined for two dissociation cuts of the water tetramer to two dimers and to a monomer plus trimer.

\section{The Energy Database}


We employed several approaches to generate a diverse database of 4-b configurations and energies.  First, configurations were obtained from direct-dynamics calculations for the water tetramer.  Second, tetramer fragments were selected from a classical dynamics simulation of water using the MB-pol potential. Third, tetramer fragments were selected from the equilibrium configuration of isomers of the water hexamer, heptamer, decamer, and 13-mer. In total, 2119 symmetry unique energies were obtained. 

The direct-dynamics calculations were done efficiently using B3LYP/6-31+G(d,p) theory. Most trajectories were initiated from the high-energy planar ring structure of the tetramer. This was generated by simply bringing the global minimum (GM) structure to a planar structure by flipping the out-of-plane H atoms to the plane of the four O atoms. This structure along with the GM one and two additional ones are given in Fig. S1 of Supplementary Material (SM). An initial kinetic energy of 1000 cm$^{-1}$ was distributed randomly in five independent trajectories to all atoms at this initial configuration. The distribution of tetramer potential energies, relative to the global minimum, is shown in Fig. S2 in SM. As seen, these range from around 2000 to nearly 12000 cm$^{-1}$. The data generated from these trajectories was pruned by simply saving every 10th configuration (in time) to obtain  468 configurations for the final database.

Additional 4-b configurations were obtained from tetramer clusters selected from a 300 K MD simulation of water using 256 monomers and the MB-pol potential. These are defined as four monomers with the third largest O-O distance smaller than 4.5 \AA. These we selected quasi-randomly from the large database of monomer configurations, and 540 configurations were selected.

For the hexamer (4 of the 8 isomers: prism, cage, book-1, and cyclic-chair), heptamer, decamer, and 13-mer water clusters, all 4-body configurations were selected at the equilibrium structure.  For the hexamer isomers the structures are from  ref. \citenum{Howard2015} and for the heptamer, decamer and 13-mer from ref. \citenum{MBE20}.

\begin{figure}[htbp!]
    \centering
    \includegraphics[width=1.0\columnwidth]{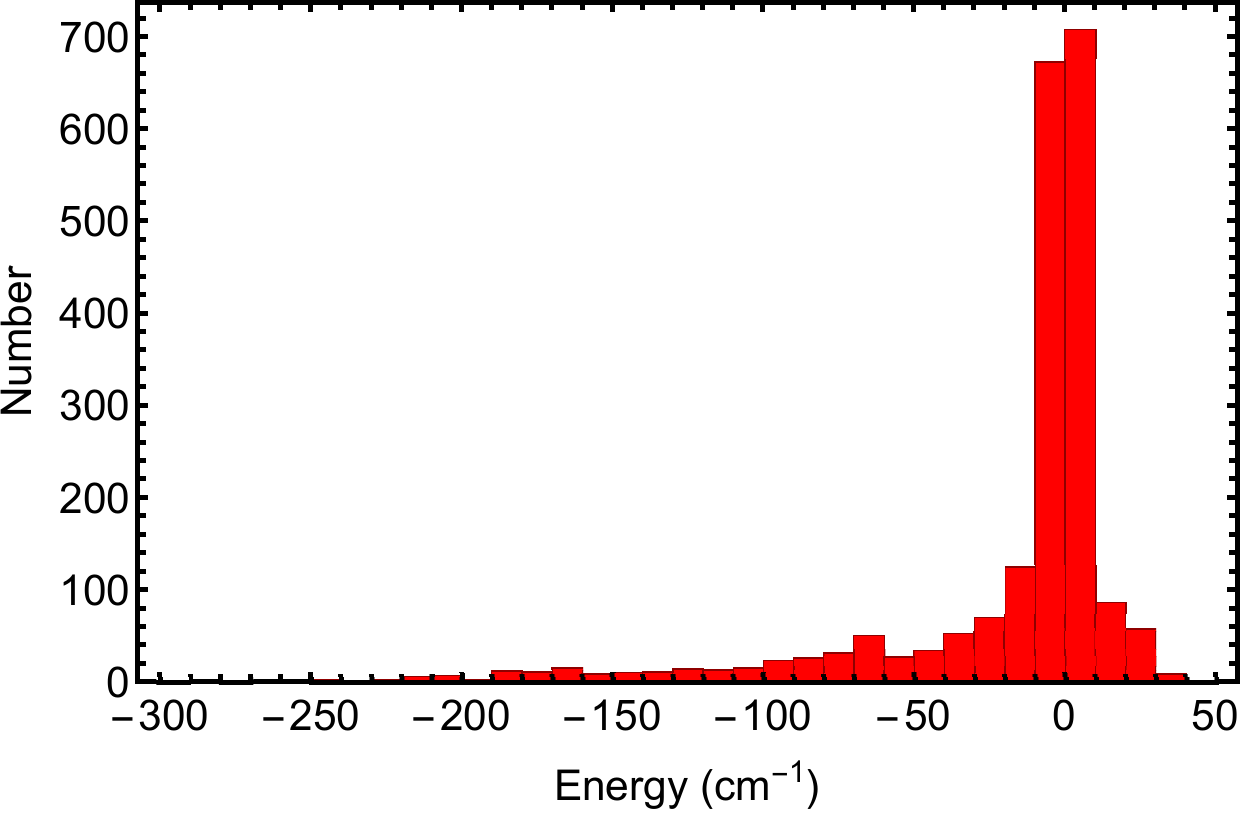}
    \caption{Histogram of the CCSD(T)-F12a/haTZ points used in the PES fit to the 4-body potential energy surface.}
    \label{fig:histof4benergies}
\end{figure}

To construct the new PES, we re-calculated the 2119 energies at CCSD(T)-F12a/haTZ level of theory. These are not corrected for basis set superposition error as this is shown to be negligible at the tetramer minimum and two other geometries, as shown in Table S1 of the SM. The entire energy range of these 4-b energies is -1063.1 to 105.2 cm$^{-1}$ and the distribution of most of these energies in a smaller range is shown in Fig. \ref{fig:histof4benergies}. There are just a small number 4-b energies outside this range.

Additional details of the these approaches are given in the SM along with the distribution of OO distances for the entire database (Fig. S5) and histograms of the 4-b energies for the heptamer and decamer (Fig. S6).

\section{The Purified PIP Basis}

The potential energy surface is fit to the data base using an expansion in permutationally invariant polynomials (PIPs).\cite{Braams2009, Xie10, conte20}  The permutational symmetry that describes all the permutations of like atoms has the 84 designation, meaning that the eight hydrogens permute with one another and the four oxygens permute with one another.  However, it is also possible to use reduced symmetries if one replicates the data set with the proper number of permutations of the water molecules.  Here we use the symmetry 22221111, meaning that the hydrogens within each water permute with one another but that they do not permute between water molecules, and that the oxygens do not permute among themselves. In order to describe the invariance with respect to the 4! permutations of water monomers we replicate the symmetry-unique configuration by 4!.  

The usual method for fitting the energy data set with the PIP polynomials is not to use the internuclear distances directly in the PIPs but rather to transform them into functions such as Morse variables ($exp(-r_{ij}/a)$) or $1/r_{ij}$ variables.  For the current application, we used a mixture of these functions -- Morse variables for the 12 intramolecular coordinates with the range parameter $a$ = 2 bohr, and $1/r_{ij}$ variables for the 54 intermolecular coordinates.

Another consideration for the basis set choice is the behavior of the PIPs as a monomer or dimer of the tetramer is removed to a great distance with respect to the remaining waters. By definition, the 4-body interaction energy must go to zero as any water monomer or water dimer is distanced.  Many of the PIPs do not have this property.\cite{Braams2009}  The process of eliminating these PIPs \cite{purified13} is what we term ``purification''.\cite{purified14, purified15c, QuConteHoustonBowman2015}  Our method of purification starts by assigning random numbers to all the internuclear distances for the four water molecules, 66 in this case, and evaluating the polynomials.  One then adds a large number to the Cartesian coordinates of the water monomers, one at a time, and water dimers, two at a time, and then calculates the new values of the polynomials.  If for any of these distancing operations the value of a polynomial does not become less than a small cutoff number (we used 10$^{-6}$), the polynomial is eliminated from the set. 
For example, for the 84 symmetry and using a maximum polynomial order of 3, there are 86 PIPs, but only 2 remain after purification.  In the 22221111 symmetry with polynomial order 3, there are 10~737 PIPs, and only 1649 remain after purification.  Of course, we want the number of PIPs, equal to the number of coefficients, to be less than the number of geometry/energy calculations, but not to be so much less that we do not obtain good fit precision.  For a data set of 2119 energy calculations, clearly 2 polynomials is insufficient.  But for a data set of 2119 $\times$ 24 = 50~856 energies, 1649 coefficients might be sufficient and accurate.  Although we considered other possibilities, as listed in Table S2, many were infeasible and, for some, the permutational replication of the data set was not effective because it mixed intra- and inter-monomer coordinates.  In the end, we used the 22221111 symmetry with a data base size of 50~856 and 1649 coefficients.

For this basis set and data base, the RMS fitting error is plotted versus the 4-b energy in Fig. S3 of the SM.  The overall RMS error for the entire dataset is 6.2 cm$^{-1}$.

\section{Results and Discussion}

To begin we present several tests of the new 4-b PES, ranging from potential cuts to a variety of water clusters.  We compare to CCSD(T)-F12a calculations as well as results from the MB-pol and TTM4-F potentials.

\subsection{Cuts of the 4-b energy}

\begin{figure}[htbp!]
    \centering
    \includegraphics[width=1.0\columnwidth]{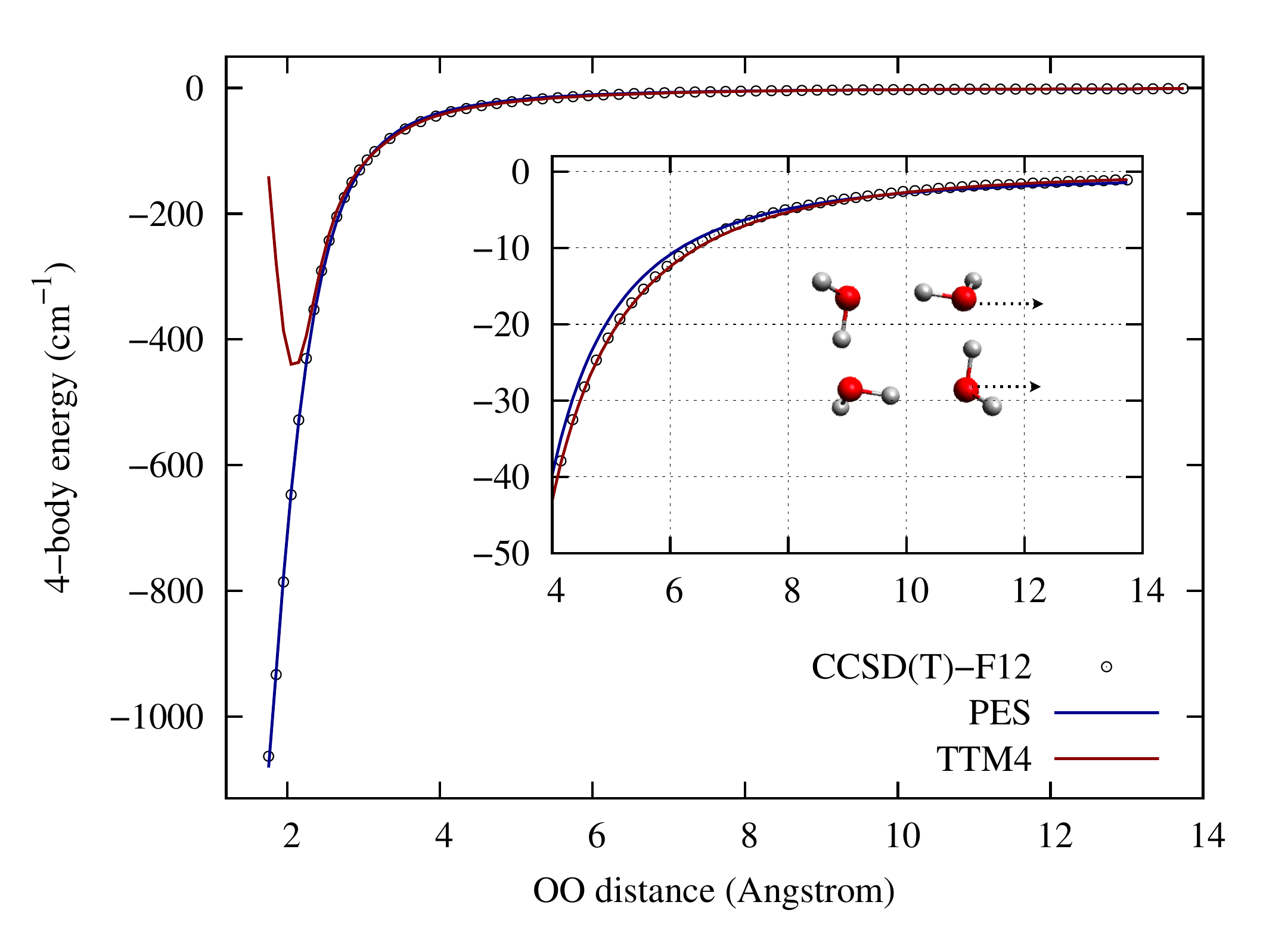}
    \caption{4-b energies from indicated sources (see text for more details) as a function the oxygen-oxygen distance between pairs of water dimers in the tetramer. The dashed arrows indicate the dimer pair that separates from the rigid tetramer. The equilibrium value of this distance is 2.7 \AA.}
    \label{fig:2+2 cut}
\end{figure}

A cut showing the 4-b energy along the separation of the water tetramer to two rigid dimers passing through the cyclic minimum (OO equal to 2.7 \AA) is shown in Figure \ref{fig:2+2 cut} from the current PES, direct CCSD(T)-F12a/haTZ, and TTM4-F calculations. Note, CCSD(T) energies for OO greater than 8 \AA are not included in the training data set.

As seen, over the range 1.7 to nearly 14 \AA, the  PES is in very good agreement with direct CCSD(T)-F12a/haTZ energies.  
The 4-b energies from the TTM4 potential are also in good agreement with the benchmarks for OO distance greater than 2 \AA.  However, a major breakdown of the accuracy of that potential is seen for shorter distances.  Evidently, this is where the classical electrostatic description of the two water dimer interaction fails, presumably due to significant chemical interaction between the electronic orbitals of the dimers. We focus here on the TTM4 potential as it plays a major role in the MB-pol water potential.\cite{mbpol3b} Namely, MB-pol is a high-level, \textit{ab initio} correction at the 2 and 3-b level to  TTM4-F in the short range and relies on that potential for all 4-b and higher body water interactions.  Finally note the gradual approach to zero in this cut.  At 11 \AA\, the energy is less than -1.0 cm$^{-1}$ 
 
A second cut showing PES and CCSD(T) energies is given in Fig. S4 in SM. Here the tetramer  dissociates to a monomer plus trimer over an OO range of 2.7 to 15 \AA, and corresponding energy range of -175 cm$^{-1}$ to 0. This is a test of PES as no CCSD(T) data for this cut were included in the fitting database.  As seen, the PES agrees well with the benchmark results and the range of the interaction for this cut is roughly 7 \AA. 
 
Both cuts show that the 4-b energy is rapidly decreasing at OO distances near and especially less than the equilibrium value of 2.7 \AA.  This indicates a significant softening of the tetramer interaction potential due to the 4-b.  This is shown in detail in Fig. S7, where a MBE decomposition of the tetramer potential is shown. Here we simply note the importance of this softening and highlight its possible significance for  high pressure conditions.

\subsection{Isomers of the water hexamer}

\begin{figure}[htbp!]
    \centering
    \includegraphics[width=1.0\columnwidth]{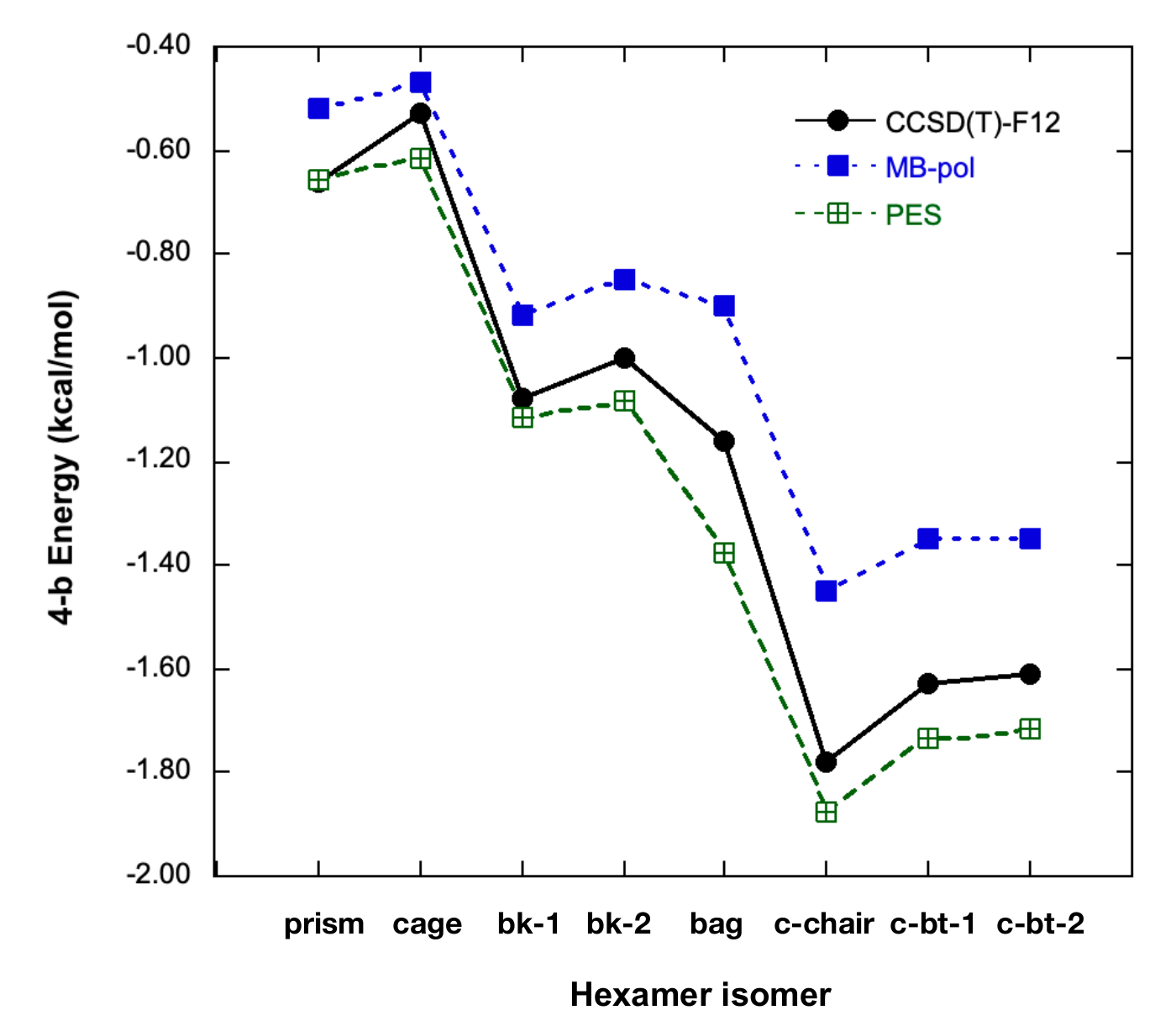}
    \caption{4-b CCSD(T) energies for indicated isomers of the water hexamer from the 4-b PES, MB-pol and previous CCSD(T)-F12/VTZ calculations.\cite{medders15}}
    \label{fig:4bhexamer}
\end{figure}

Another test of the PES is the 4-b energies of the isomers of the water hexamer, prism, cage, book-1 (bk-1), book-2 (bk-2), bag, cyclic-chair (c-chair), cyclic-boat-1 (c-bt-1) and cyclic-boat-2 (c-bt-2).  These isomers have been the focus of a number of papers, and the one by Medders et al. is of particular interest.\cite{medders15} There the 4-body energy from several water potentials, including WHBB5 and MB-pol were compared with  CCSD(T)-F12/VTZ energies for eight isomers. Here we focus on the errors in the MB-pol and the new 4-b PES.  These are shown in Fig.\ref{fig:4bhexamer}.  As seen, the 4-b PES errors are significantly smaller than those from MB-pol which have been discussed above and also in the literature. A numerical comparison of 4-b energies from other potentials as well are given in Table S3 in the SM. (It should be noted that some of the hexamer energies were included in the training dataset and so the good agreement between the PES and \textit{ab initio} energies is not unexpected.) 

\subsection{Larger water n-mers: n= 7, 10, 13, 16, 21}
One aspect of an accurate 4-b PES is that it can be used for arbitrarily many monomers with virtually zero BSSE.  Thus, it is  possible  to test the prediction of  4-b energies of larger water clusters using the 4-b PES and compare the results to benchmark calculations. A convenient set of benchmark calculations has been published by Heindel and Xantheas,\cite{MBE20} who provided geometries (in supporting information) and reported many-body energies calculated with and without BSSE-correction at several levels of theory.  Table \ref{tab:tab_nmers} shows their results in columns 4 and 5 for MP2 and the basis set listed in column 3. The results using the PES to calculate all 4-b energy in each n-mer are shown in the last column.  The agreement is good through $n$ = 16. Remarkably, even better agreement is obtained between the PES and benchmark CCSD(T)-F12a/haTZ calculations performed up to the 13-mer. For the two isomers of $n$ = 21 the BSSE correction to MP2 calculations was not applied and coupled clusters energies were not obtained as both computations were evidently too intensive. As seen, the PES 4-b results are much less than the uncorrected MP2/aVDZ results.  While this is certainly qualitatively correct, the 4-b PES results are perhaps just semi-quantitatively correct for these larger clusters.  
 
\begin{table}
\begin{threeparttable}[htbp!]
\centering
\caption{4-b interaction energies (kcal/mol) for various water clusters, where the information in columns 3--5 is taken from the SI in ref. \citenum{MBE20}, numbers in column 6 are from our CCSD(T)-F12/haTZ calculations, and the numbers in column 7 are calculated using our 4-b PES.}

\begin{tabular}{|c|c|c|c|c|c|c|}
\hline
\ce{(H2O)_n} & No. of  & H-X   &     &    MP2    & CCSD(T)-   & 4-b \\
     n=      & 4mers   & basis & MP2 &   BSSE-   & F12/haTZ & PES \\
             & in nmer &       &     & corrected &  &     \\
\hline
  7\tnote{a} &   35 & aVTZ &  -1.106 & -0.874 & -0.987 & -1.012 \\
 10\tnote{a} &  210 & aVTZ &  -3.028 & -1.978 & -2.576 & -2.341 \\
 13\tnote{a} &  715 & aVDZ &  -6.352 & -1.499 & -1.539 & -1.571 \\
 16\tnote{a} & 1820 & aVDZ &  -9.463 & -2.179 &    -   & -2.983 \\
 21\tnote{b} & 5985 & aVDZ & -20.976 &    -   &    -   & -9.888 \\
 21\tnote{c} & 5985 & aVDZ & -19.892 &    -   &    -   & -5.920 \\
\hline
\end{tabular}

\begin{tablenotes}
\item[a] These are global minimum structures;
\item[b] Fully solvated structure;
\item[c] All surface structure. 
\end{tablenotes}

\label{tab:tab_nmers}
\end{threeparttable}
\end{table}

\begin{figure}[htbp!]
    \centering
    \includegraphics[width=1.0\columnwidth]{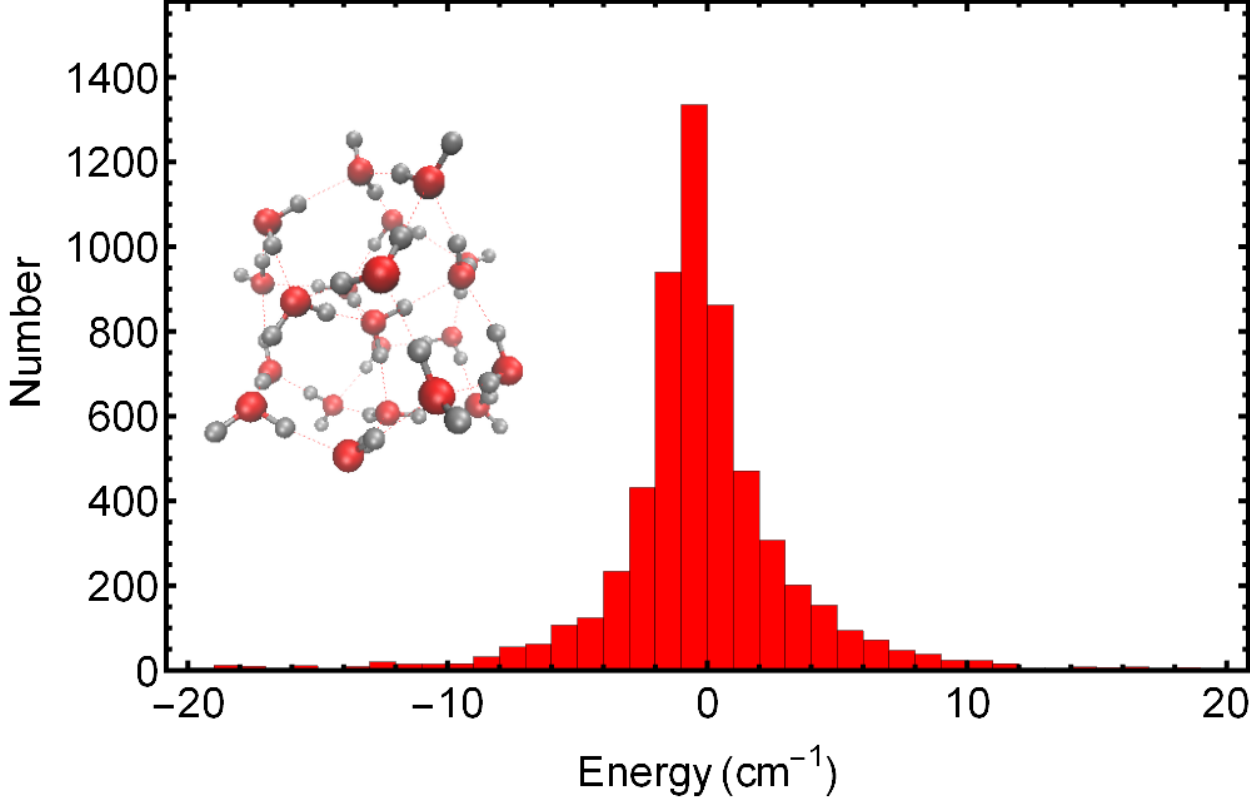}
    \caption{Distribution of PES 4-b energies for the fully solvated 21-mer}
    \label{fig:Histo_21}
\end{figure}

For most of the large clusters, many of the 4-mer configurations involve at least one water that is quite distant from the others, so that the 4-b energy is quite small.  For example, Fig. \ref{fig:Histo_21} shows a histogram of the PES-determined 4-b energies for the ``fully-solvated'' 21-mer.  There are large peaks near zero, although the distribution actually stretches from -95 to 54 cm$^{-1}$, with only a minor number of energies outside the range depicted in the Figure.  Since the total 4-b energy is the sum of all the 4-body contributions (5985 in the case of $n$ = 21), there is evidently a large degree of cancellation.

\subsection{Timings}
Finally, an important aspect of the new 4-b PES is the computational effort to use it. A single evaluation of the 4-b PES takes 2.45 seconds for 100 000 potential calls on a single core of the 2.4 GHz Intel Xeon processor. This is significantly smaller than the computational effort to evaluate our WHBB PES for the water trimer.   

While this computational effort is not large, it should be recalled that the number of 4-b interactions go formally as  $N^4$ for $N$ water monomers, Thus, the cost would be high if all the tetramers were to be evaluated.  Several strategies could be used to reduce the cost for ``condensed phase'' simulations. One approach is to use a distance-based cutoff via a standard switching function. This would greatly reduce the number of tetramers to be evaluated because in large clusters many tetramers have large inter-monomer distances and the 4-b energy for such tetramers is essentially zero and thus not necessary to evaluate. For example, a cluster with 256 monomers, if we only evaluate the potential of tetramers whose largest O-O distance is smaller than 9 \AA, 99.7\% of the tetramers are dropped, and calculating the total 4-b energy takes 14 seconds rather than 1 hour. Parallelization is another possible strategy since the calculation of one tetramer is independent of the others. We plan to investigate the efficiency of the 4-b PES in the future.

\section{Conclusions}

We reported the first \textit{ab initio}, full-dimensional potential surface for the 4-body interaction of water. The potential is a purified PIP fit to a diverse set of 2119 unique CCSD(T)-F12a/haTZ energies. This representation ensures that the PES rigorously goes to zero as a monomer or dimer is separated from a tetramer cluster.  Tests of the 4-b PES demonstrate the high-fidelity compared to benchmark calculations.  In addition, the 4-b PES was shown to significantly reduce errors in 4-b interaction energies for isomers of the water hexamer obtained from a high-level water potential that describes water interactions to the 3-b level. This new 4-b PES was shown to be fast to evaluate while for applications to large numbers of water monomers a large fraction of 4-b configurations are beyond the range of the 4-b PES and so do not have to be evaluated. This was demonstrated for the 21-mer.

The current 4-b PES could be an add-on to a water potential that either does not contain 4-b terms or have approximate treatments of the 4-b, e.g., the TTMn family of potentials.  

Finally, it should be clear that the present 4-b PES is ``version 1.0".  And while it appears to be both fast to evaluate and accurate, we anticipate being able to make significant speed-ups, the easiest one of which is to make use of multi-core architecture of all modern workstations and computer nodes.

\section*{Supplementary Material}

The supplementary material contains additional details of the database of energies, the purified PIP basis, an analysis of the PES fitting RMS error, a plot of the 4-b energy for the dissociation of the tetramer to the trimer plus monomer from the PES and CCSD(T)-F12a energies, a table showing 4-b energies for isomers of the water hexamer from the present CCSD(T)-F12a/haTZ calculations, the PES and previously reported TTM4-F, and MB-pol potentials,\cite{medders15} and histograms of the distribution of OO distances for database and also for selected water clusters up to the 21-mer.


\section*{Acknowledgment}
JMB thanks NASA (80NSSC20K0360) for financial support. We thank Francesco Paesani for sending the 300 K classical trajectory data for 256 waters using the MB-pol potential.\\

\section*{Data Availability}
The data that support the findings of this study are available from the corresponding authors upon reasonable request. The new 4-b PES for water is provided as supplementary material.

\bibliography{refs.bib}
\end{document}


\title{Supplementary Material: A CCSD(T)-based permutationally invariant polynomial 4-body potential for water}
\date{\today}
\author{Apurba Nandi}
\email{apurba.nandi@emory.edu}
\affiliation{Department of Chemistry and Cherry L. Emerson Center for Scientific Computation, Emory University, Atlanta, Georgia 30322, U.S.A.}
\author{Chen Qu}
\affiliation{Department of Chemistry \& Biochemistry, University of Maryland, College Park, Maryland 20742, U.S.A.}
\author{Paul L. Houston}
\email{plh2@cornell.edu}
\affiliation{Department of Chemistry and Chemical Biology, Cornell University, Ithaca, New York
14853, U.S.A. and Department of Chemistry and Biochemistry, Georgia Institute of
Technology, Atlanta, Georgia 30332, U.S.A}
\author{Riccardo Conte}
\email{riccardo.conte1@unimi.it}
\affiliation{Dipartimento di Chimica, Universit\`{a} Degli Studi di Milano, via Golgi 19, 20133 Milano, Italy}
\author{Joel M. Bowman}
\email{jmbowma@emory.edu}
\affiliation{Department of Chemistry and Cherry L. Emerson Center for Scientific Computation, Emory University, Atlanta, Georgia 30322, U.S.A.}

\maketitle

The supplementary material contains an examination of BSSE for CCSD(T) calculations of the 4-b energies of the water tetramer,  details of the purified PIP basis, an analysis of the PES fitting RMS error, a test for the dissociation along a cut leading to a monomer+trimer, a table showing 4-b energies for isomers of the water hexamer from the present CCSD(T)-F12a/haTZ calculations, the PES and previously reported TTM4-F, MB-Pol potentials, distributions of OO distances for the database of configuration and distribution of 4-b energies for the heptamer and decamer water clusters and a MBE decomposition of the tetramer potential for the dimer-dimer cut.

\subsection*{Basis Set Superposition Error Analysis}
\begin{figure}[htbp!]
    \centering
    \includegraphics[width=0.8\columnwidth]{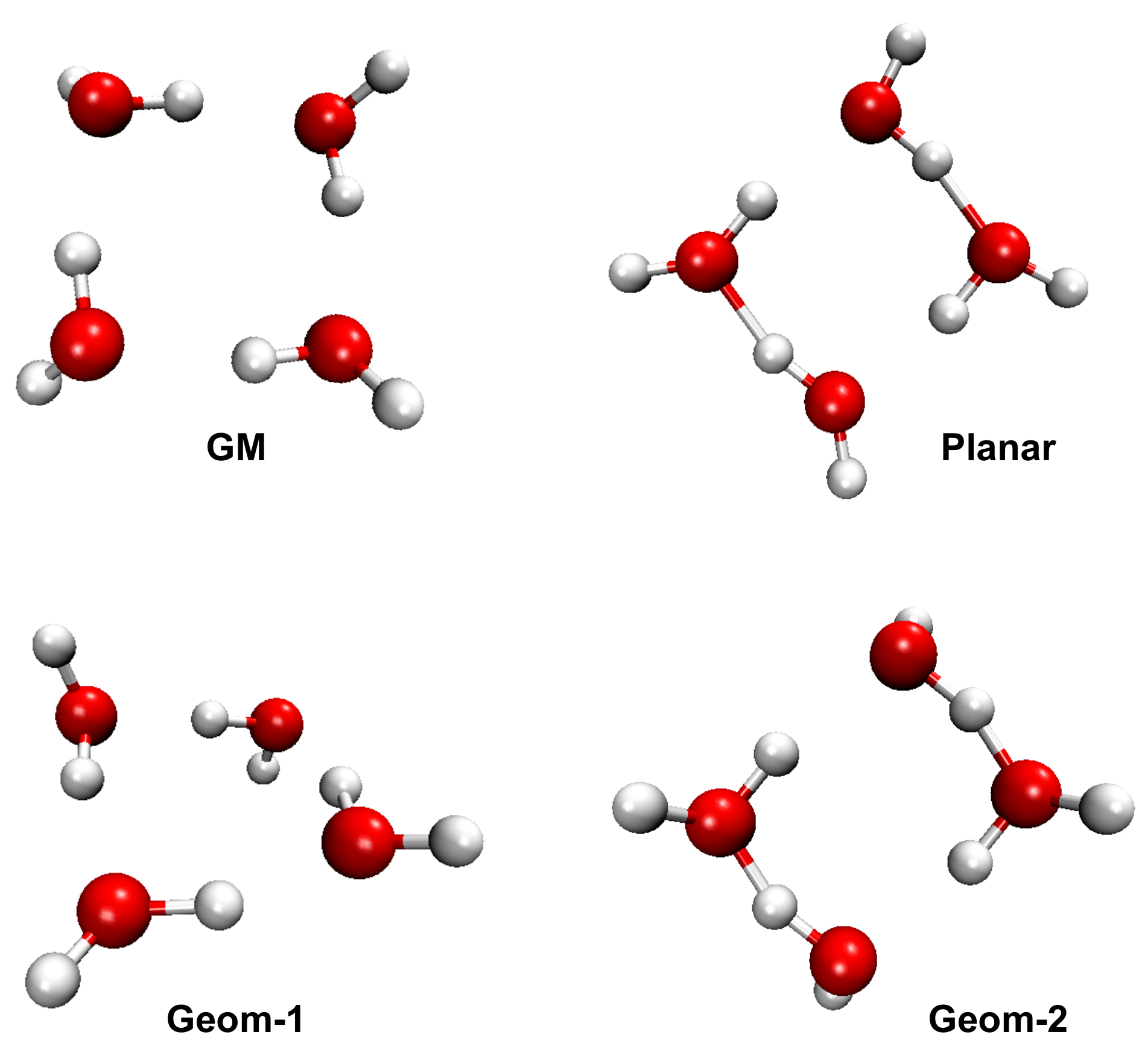}
    \caption{Water tetramer Geometries.}
    \label{fig:tetGEOM}
\end{figure}

We considered four configurations for the water tetramer, shown in Fig. \ref{fig:tetGEOM}

\begin{table}[htbp!]
\centering
\caption{Comparison of BSSE corrected and uncorrected 4-b interaction energies (in cm$^{-1}$) for minimum geometry as well as two more geometries (Geom-1 and Geom-2) at various level of theory.}
\label{tab:tab_ch4_2}

	\begin{tabular*}{1.0\columnwidth}{@{\extracolsep{\fill}}ccccccc}
	\hline
	\hline\noalign{\smallskip}
	  & \multicolumn{2}{c}{Minimum} & \multicolumn{2}{c}{Geom-1} & \multicolumn{2}{c}{Geom-2} \\
    \noalign{\smallskip} \cline{2-3} \cline{4-5} \cline{6-7} \noalign{\smallskip}
     Method & corrected & uncorrected & corrected & uncorrected & corrected & uncorrected \\
	\noalign{\smallskip}\hline\noalign{\smallskip}
     CCSD(T)-F12a/haDZ & -173.1 & -169.4 & -402.6 & -395.3 & -636.2 & -636.0 \\
     CCSD(T)-F12a/haTZ & -174.4 & -174.5 & -405.6 & -404.0 & -647.2 & -646.2 \\
     CCSD(T)-F12a/haQZ & -175.9 & -175.3 & -406.8 & -404.8 & -649.4 & -648.7 \\
	\noalign{\smallskip}\hline
	\hline
   
	\end{tabular*}

\end{table}

\subsection*{Database generation}
Several strategies were employed to generate the data sets. First one is by performing \textit{ab initio} molecular dynamics (AIMD) simulations at several total energies by using microcanonical sampling (NVE). Initial conditions were chosen to obtain a wide coverage of the configuration space. These AIMD trajectories were propagated for 30000 time steps with the step size of 5.0 a.u. (about 0.12 fs) and with total energies of 8000, and 11000 cm$^{-1}$. The geometries were recorded every 30 time
steps from each trajectory to generate the data set and the final data set consist of 4000 energies recorded from four different trajectories. These calculations were done at the efficient DFT(B3LYP)/6-31+G(d,p) level of theory, using the Molpro quantum chemistry package.\cite{MOLPRO_brief} The distributions of the electronic energies are shown below.

\begin{figure}[htbp!]
    \centering
    \includegraphics[width=0.8\columnwidth]{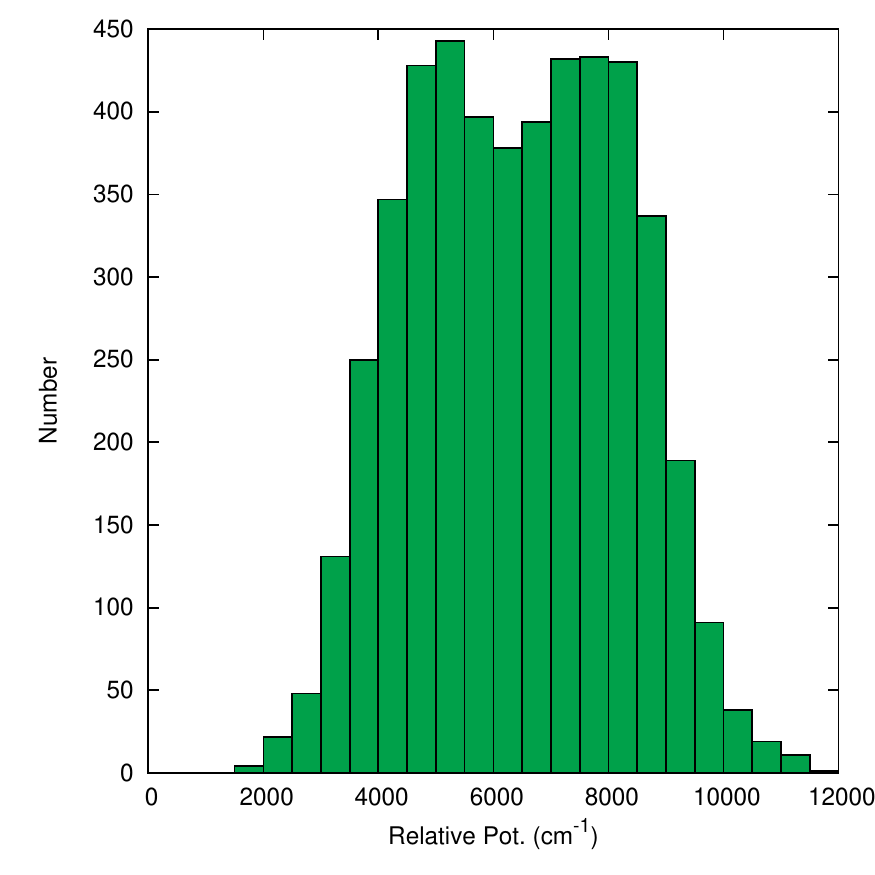}
    \caption{Tetramer potential energies from direct dynamics trajectories}
    \label{fig:tetAIMD}
\end{figure}

Additional configurations were obtained from tetramer clusters selected from a 300 K MD simulation of water using 256 monomers and the MB-pol potential. From each snapshot of the MD trajectory, a point in the interior of the cluster was randomly selected and a sub-cluster formed by water molecules that are within 5 \AA\, to the random point was extracted. Then all the tetramers from the sub-cluster whose third largest O-O distance is smaller than 4.5 \AA\, were extracted. Similar tetramer structures were then removed, and finally 540 configurations were randomly sampled from the tetramer geometries that remain.

Sixty one energies were obtained from the cut shown in the main text dissociating to the two dimers over the range 1.7 to 7.7 \AA.  

For the hexamer (4 of the 8 isomers: prism, cage, book-1, and cyclic-chair), heptamer, decamer, and 13-mer water clusters, all 4-body configurations were selected at the equilibrium structure.  For the hexamer isomers the structures are from  ref.\citenum{Howard2015} and for the heptamer, decamer and 13-mer from ref. \citenum{MBE20}.

\subsection*{Purified basis and precision analysis}
The purified PIP basis is one in which all permutationally invariant polynomials go to zero in relevant asymptotic regions.  For the present application to the 4b water potentials these are  one mononer plus one trimer and two dimers.  There are 4 possible monomer + trimer arrangements and 6 possible dimer+dimer arrangements and the purified PIP basis describes all of these.

\begin{table}[htbp!]
\caption{Symmetry and PIP information for water tetramer clusters}

\begin{tabular*}{1.0\columnwidth}{@{\extracolsep{\fill}}cccccc}
\hline
\hline\noalign{\smallskip}
 &  &  &  & poly & required \\
 sym\_order & nmers & mono & poly & after  & permuted  \\
  &  &  &  & purif. & data sets \\
 \noalign{\smallskip}\hline\noalign{\smallskip}
 \text{22221111$\_$4} & \text{1+1+1+1} & 20581 & 141765 & 51444  & 24 \\
 \text{22221111$\_$3} & \text{1+1+1+1} & 2911 & 10737 & 1649  & 24 \\
 \text{4422$\_$3} & \text{2+2} & 17019 & 896 & 18  & 6$^a$ \\
 \text{4422$\_$4} & \text{2+2} & 233246 & 7173 & -  & 6$^a$ \\
 \text{6321$\_$3} & \text{3+1} & 21230 & 636 & 18 & 4$^a$ \\
 \text{84$\_$3} & 1 & 35708 & 86 & 2 &  1 \\
 \text{84$\_$2} & 1 & \text{msa fails} & \text{} & \text{}  & 1 \\
\noalign{\smallskip}\hline
\hline
\end{tabular*}
\begin{tablenotes}
\item[a] a. These replications do not achieve permutational symmetry among the waters because they interchange inter- and intra-monomer distances.
\end{tablenotes}
\label{tab:tab_nmers}
\end{table}

\begin{figure}[htbp!]
\centering
    \includegraphics[width=0.5\columnwidth]{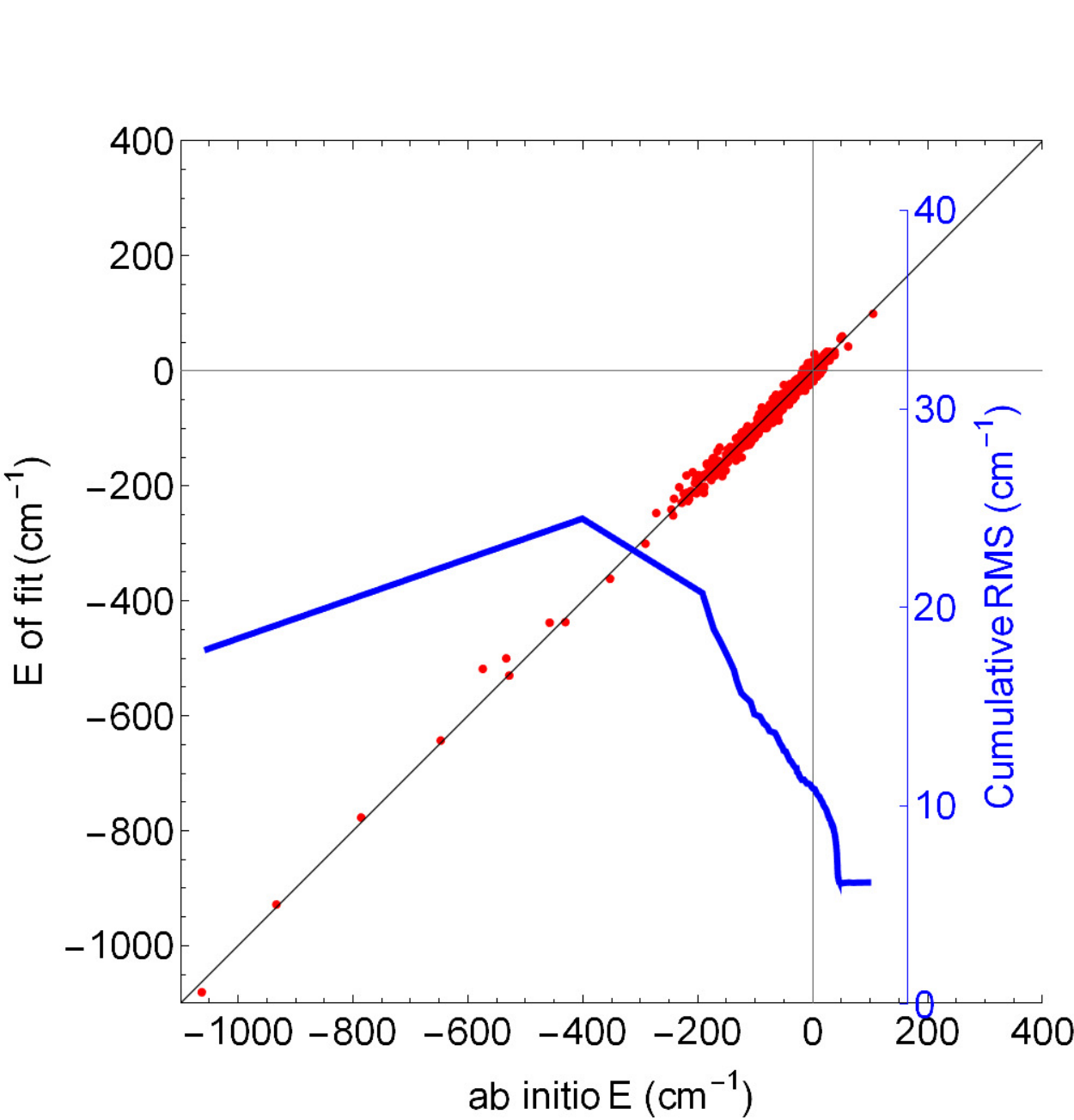}
    \caption{Correlation plot between \textit{ab initio} energies and energies calculated by the PES fit. The correlation coefficient is R$^2$ = 0.990 and the RMS error is 6.2 cm$^{-1}$. (blue) The cumulative RMS error in cm$^{-1}$. }
    \label{fig:correlation}
\end{figure}


\begin{figure}[htbp!]
\centering
    \includegraphics[width=0.7\columnwidth]{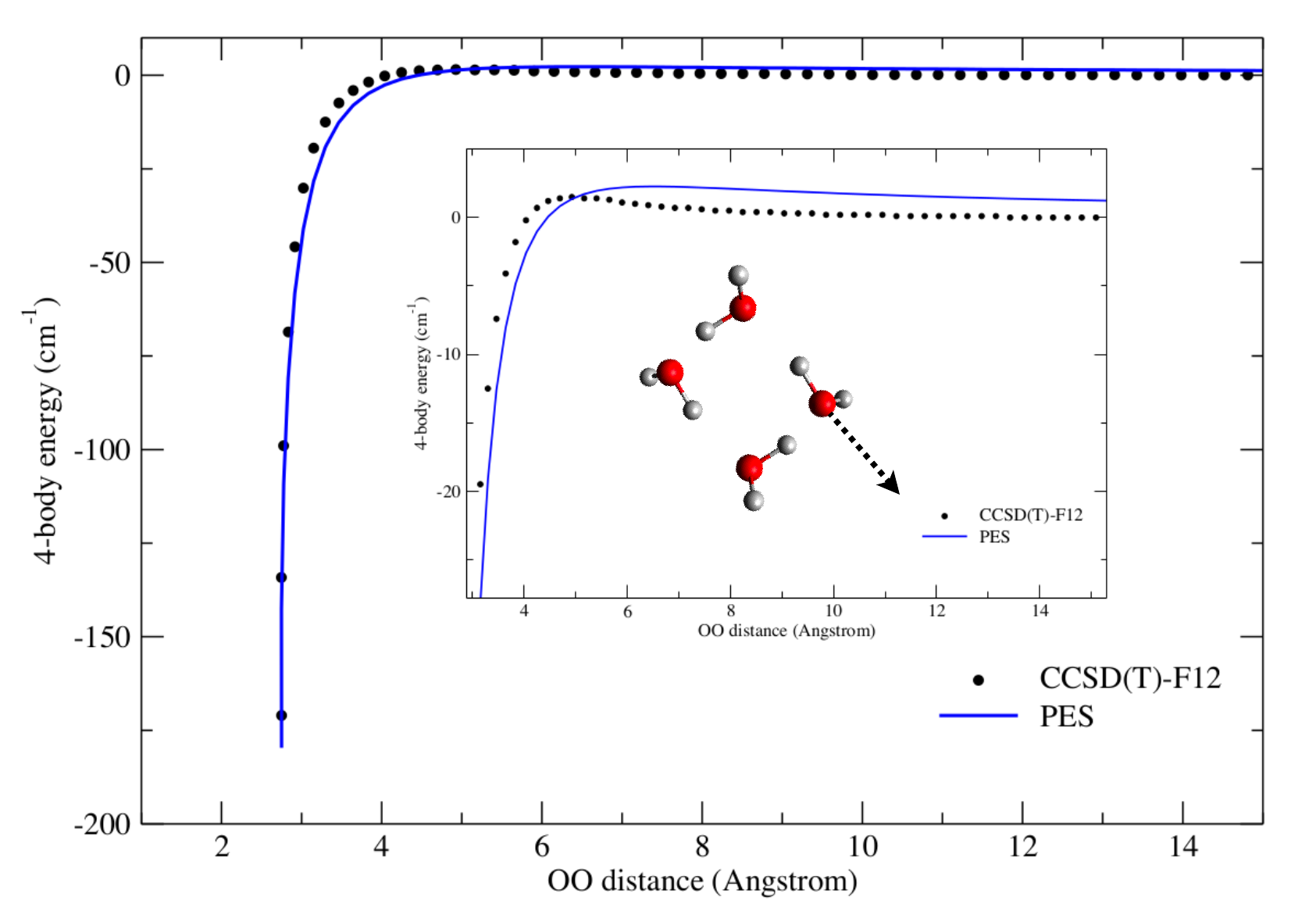}
\caption{ A test of the 4-b PES for a cut with a single monomer separating from the tetramer against CCSD(T)-F12a/haTZ energies. OO is the distance between the O atoms on the two monomers on the axis inferred from the arrow.}
    \label{fig:1+3cut}
\end{figure}

\begin{figure}[htbp!]
\centering
    \includegraphics[width=1.0\columnwidth]{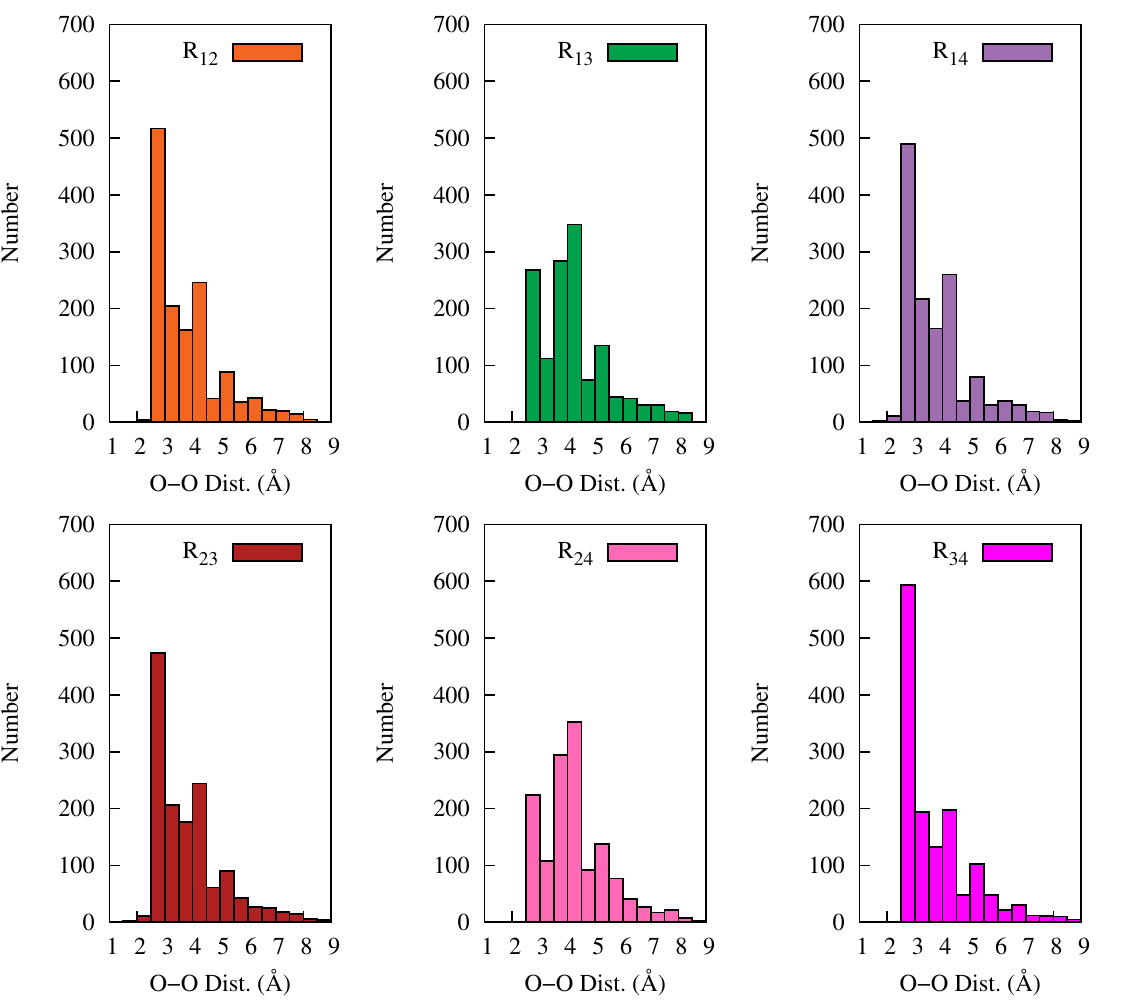}
\caption{Distribution of OO distances for the entire dataset.}
    \label{fig:histofOO}
\end{figure}

\begin{figure}[ht]
    \centering
    \includegraphics[width=0.8\columnwidth]{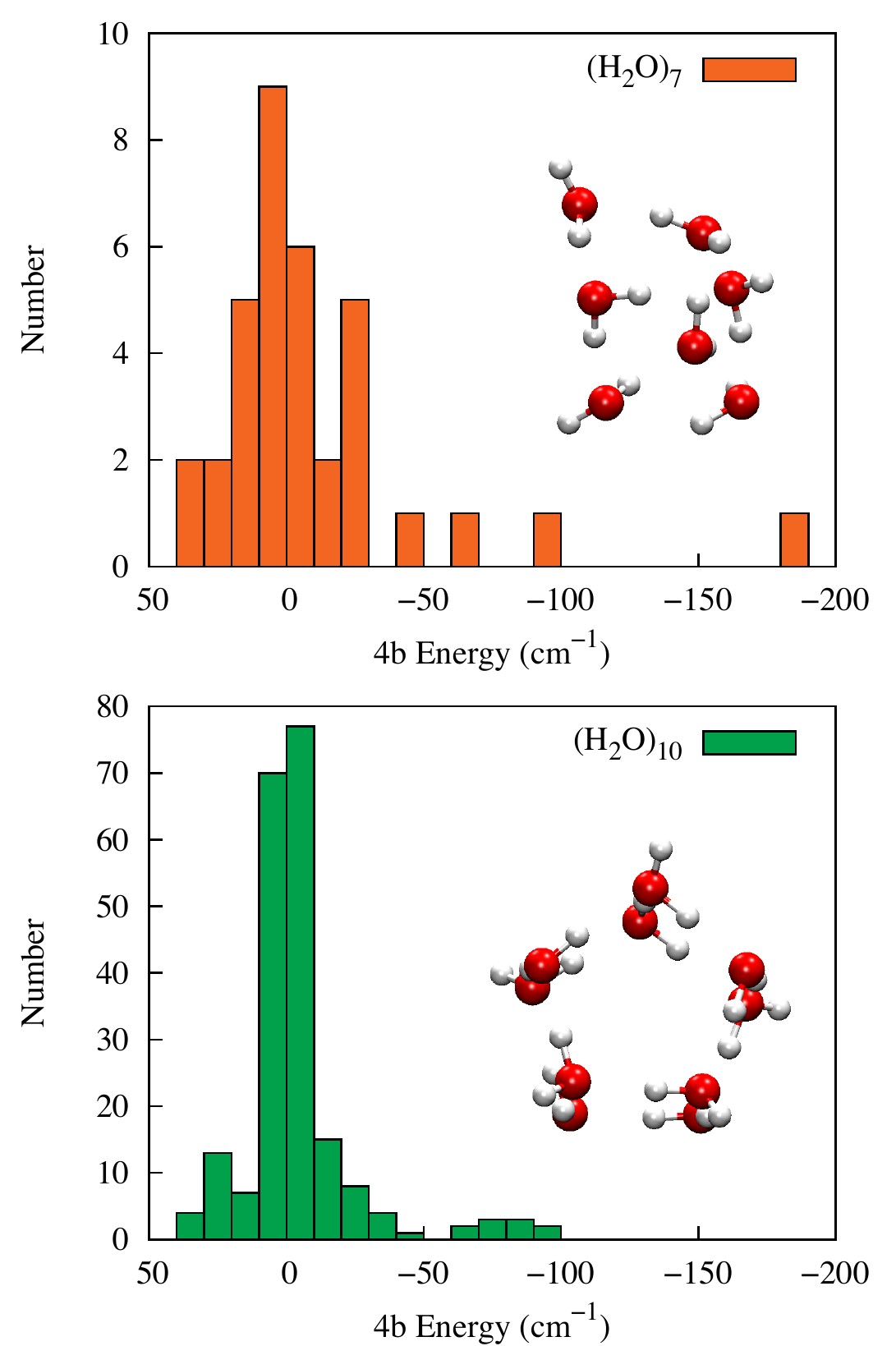}
    \caption{Histograms of 4-b energies for heptamer and decamer.}
    \label{fig:4b_Hist}
\end{figure}

\begin{figure}[htbp!]
    \centering
    \includegraphics[width=0.8\columnwidth]{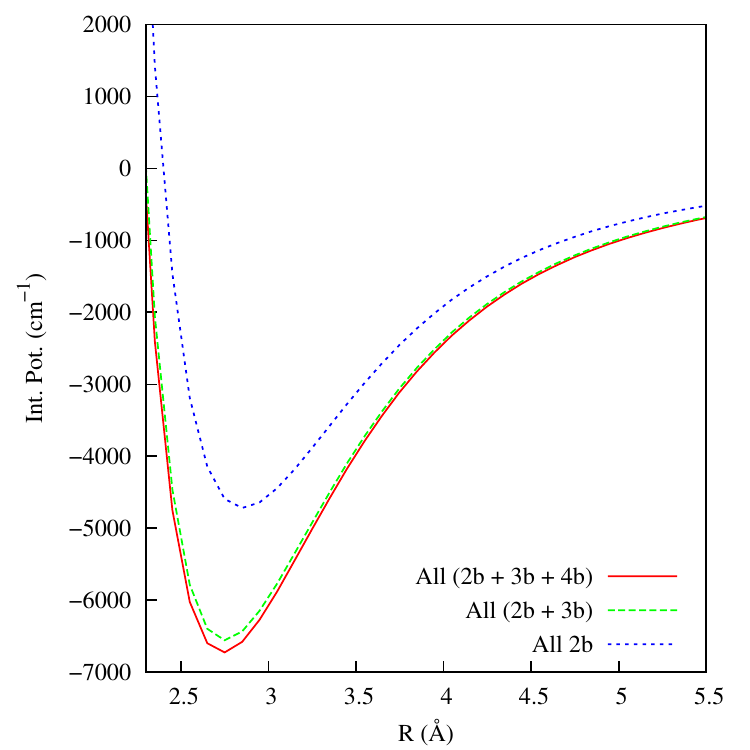}
    \caption{Decomposition of the tetramer potential as indicated. The interaction potentials are computed at MP2/aug-cc-pVTZ level of theory along the dimer-dimer cut. Note there is only one 4-b interaction.}
    \label{fig:Sum_Int_Pot}
\end{figure}

\begin{table}[htbp!]
\centering
\caption{4-b interaction energies (in kcal/mol) of eight isomers of the water hexamer from indicated sources.}
\label{tab:tab_2}

\begin{threeparttable}[htbp!]
	\begin{tabular*}{\columnwidth}{@{\extracolsep{\fill}}lcccc}
	\hline
	\hline\noalign{\smallskip}
     Isomer & PES & Present CCSD(T)-  & CCSD(T)       & MB-pol  \\
            &     & F12a/haTZ      & -F12/VTZ\tnote{a} & /TTM4-F \\
	\noalign{\smallskip}\hline\noalign{\smallskip}
     Prism  & -0.66 & -0.62 & -0.66 & -0.52 \\
     Cage   & -0.61 & -0.54 & -0.53 & -0.47 \\
     Book-1 & -1.12 & -1.16 & -1.08 & -0.92 \\
     Book-2 & -1.08 & -1.05 & -1.00 & -0.85 \\
     Bag    & -1.38 & -1.21 & -1.16 & -0.90 \\
     Chair  & -1.88 & -1.85 & -1.78 & -1.44 \\
     Boat-1 & -1.73 & -1.71 & -1.63 & -1.35 \\
     Boat-2 & -1.71 & -1.69 & -1.61 & -1.35 \\
	\hline
   
	\end{tabular*}

   	\begin{tablenotes}
    \item[a] From Medders et al J. Chem. Phys. 143, 104102 (2015).
    \end{tablenotes}	
\end{threeparttable}

\end{table}



 
   
    

\clearpage
\bibliography{refs.bib}